\documentclass[twocolumn,showpacs,preprintnumbers,amsmath,amssymb]{revtex4}
\usepackage{graphicx}

\newcommand{\ie}{{\it i.e.}}

\def\bra#1{ \langle #1\!\mid }
\def\ket#1{\mid\!#1\rangle}

\begin{document}

\title{Atom-to-molecule conversion efficiency and adiabatic fidelity}

\author{Li-Hua Lu and You-Quan Li}
\affiliation{Zhejiang Institute of Modern Physics and Department of Physics,\\
Zhejiang University, Hangzhou 310027, P. R. China}

\begin{abstract}
The efficiency of converting two-species fermionic atoms into bosonic
molecules is investigated in terms of mean-field Lagrangian density.
We find that the STIRAP technique aided by Feshbach resonance
is more effective than the bare Fechbach resonance for $^6$Li atoms
rather than $^{40}$K atoms.
We also make general consideration on the symmetry and its relevant
conservation law, which enable us to introduce a natural
definition of adiabatic fidelity for CPT state.
The calculated values of the fidelity then provide an interpretation
on why the conversion efficiencies for $^{40}$K and $^6$Li
are distinctly different.
\end{abstract}
\received{6 December 2007}
\maketitle

\section{Introduction}

There has been much attention to the molecular Bose-Einstein
condensate (BEC) which is a versatile playground not only for cold
atomic physics experiments but also for other research areas, such
as condensed matter theory as well as quantum field theory. It is
no longer a pure Bose system as the molecular BEC can be a system of
Bose-Fermi mixtures. This makes the theoretical structure rich and
colorful. To realize the molecular BEC, one must create stable
molecules with long lifetime. In the recent
experiments~\cite{Greiner,Joch,zwi}, the technique of Feshbach
resonance plays an important role in the creation of molecules.
Since the molecules created through such a technique usually
suffer from fast decay due to the vibrational excitation,
the stimulated Raman adiabatic passage (STIRAP) in
photoassociation~\cite{Matt} has been regarded as an effective
approach to create ground-state molecules.
The success of STIRAP technique requires
the existence of a coherent population trapping (CPT) state which can be followed
adiabatically~\cite{Alze}.
Such a condition can be fulfilled  for
linear $\Lambda$ systems by appropriately choosing  laser
frequencies. However, for the system with inter-particle
interactions, the two-photon resonance condition dynamically
changes when population is transferred from atomic states to
molecular states.
This makes the CPT state more difficult to be
followed adiabatically.
The adiabatic property was studied by
means of adiabatic fidelity in a recent theoretical
work~\cite{Ying} for a simplified model
of monoatomic system without inter-particle interaction.
It is worthwhile to appropriately define the adiabatic fidelity to
study the adiabatic property for a more realistic system,
such as the diatomic system which has been realized
in serval experiments~\cite{Regal,Stre,Stan, Ino,JC,Jin}.

In this paper, we consider systems consisting of fermionic atoms
in different hyperfine states and their compounded molecules
coupled through the STIRAP technique aided by Feshbach resonance.
We discuss systems of $^6$Li and $^{40}$K as concrete examples.
In section~\ref{sec:general}, we model the systems
with inter-particle interaction and derive the dynamical equations
for them.
We make a general consideration on the symmetry and
the relevant conservation law, and then introduce the definition
of fidelity for our system.
In section~\ref{sec:cpt-state}, we look for solutions of
the CPT state.
In section~\ref{sec:efficiency}, we calculate the conversion efficiency
for $^{40}$K atoms and $^6$Li atoms, respectively, and discuss the
corresponding features.
In section~\ref{sec:fidelity}, we
study the relation between the atom-to-molecule conversion
efficiency and the adiabatic fidelity for CPT state.
We also study the effect of the decay of quansibound molecules
and compare the difference between
$^{40}$K  and $^6$Li systems.
Our results are briefly summarized in section~\ref{sec:summary}.

\section{Model and its general properties}\label{sec:general}

We consider that two species of fermionic atoms
are converted into stable molecules via the STIRAP technique
aided by Feshbach resonance.
Here we use $\ket{a}$ and $\ket{b}$ stand for the states of the
free atoms in the open channel,
and $\ket{m}$ and $\ket{g}$ for the quasibound and ground
molecular states in the close channel, respectively.
A pair of ferimonic atoms are coupled with the quasibound
molecular state $\ket{m}$ through  a Feshbach resonance, where
the coupling strength is denoted by $\alpha'$ and the detuning by
$\mathcal{E}'$.
Additionally, the sates $\ket{m}$ and $\ket{g}$
are coupled with each other through a laser field with the
coupling strength $\Omega'$ and detuning $\Delta'$.
Since the system is a Fermi-Bose mixture whose Hilbert space
actually carries out the representation of a graded unitary group
SU$(2\mid2)$,
it will be an arduous work to study the dynamics of
the system by means of the Heisenberg equation of motion.
Whereas, with the help of mean-field Langrange density~\cite{Lu},
one can investigate the dynamics conveniently,
\begin{eqnarray}\label{eq:Lag}
L\!\!\!&&=\sum_i \Bigl( \frac{i\hbar}{2}
  \bigl(\psi_i^*\frac{\partial\psi^{}_i}{\partial t}
  -\psi^{}_i\frac{\partial\psi_i^*}{\partial t}\bigr)-T_i \Bigr)
  -\mathcal{E}'\psi_m^*\psi_m^{}
          \nonumber\\
&&-\alpha'\bigl(\psi_m^*\psi^{}_a\psi^{}_b+\mathrm{H.c.}\bigr)
  +\Omega'\bigl(\psi_m^*\psi_g +\mathrm{H.c.}\bigr)
     -\Delta'\psi_g^*\psi_g
            \nonumber\\
&&-\frac{1}{2}\sum_{i\neq
j}\lambda'_{ij}|\psi^{}_i|^2|\psi^{}_j|^2
-\frac{1}{2}\Bigl(\lambda_{mm}'|\psi^{}_m|^4 +\lambda_{gg}'|\psi^{}_g|^4\Bigr)
            \nonumber\\
&&-\frac{3}{5}\bigl(A_a'|\psi^{}_a|^{10/3}+A_b'|\psi_b^{}|^{10/3}\bigr),
\end{eqnarray}
where $T_i$ denotes the kinetic energy term of the  $i$th
component and $i,j=a, b, m, g$.
Here the coefficients
$\lambda'_{ij}=\lambda'_{ji}=2\pi\hbar^2 a_{ij}^{}/m^{}_{ij}$ are
the interaction strengths between particles with $m_{ij}^{}=m^{}_i
m^{}_j/(m^{}_i+m^{}_j)$ being the reduced mass and $a^{}_{ij}$ the
$s$-wave scattering length.

\subsection{The conservation law}

The Lagrangian (\ref{eq:Lag}) is no more invariant under a simultaneous global
phase transformation due to the presence of the atom-to-molecule conversion term.
As the Lagrangian does not include the term flipping the two fermionic
components into each other, their corresponding phase parameters
are not necessarily the same in general.
One can find that the Lagrangian Eq.~(\ref{eq:Lag})
is invariant under the phase transformation,
$(\psi_a, \psi_b, \psi_m, \psi_g)\mapsto
  (\psi_a, \psi_b, \psi_m, \psi_g)U(\vartheta_a,\vartheta^{}_b )$
with
\begin{eqnarray}\label{eq:trans}
U(\vartheta_a,\vartheta^{}_b )
&=&
\left(%
\begin{array}{cccc}
  e^{i\vartheta_a} & 0 & 0 & 0\\
  0 & e^{i\vartheta^{}_b} & 0 & 0 \\
  0 & 0 & e^{i(\vartheta_a +\vartheta^{}_b )} & 0 \\
  0 & 0 & 0 & e^{i(\vartheta_a +\vartheta^{}_b )} \\
\end{array}%
\right).
\end{eqnarray}

Following the formulism of Noether theorem,
we evaluate the variation of the action caused by the infinitesimal
phase transformation $U(\delta\vartheta_a,\delta\vartheta^{}_b )$,
\begin{eqnarray*}\label{eq:general-variation}
\delta I = \int
  \bigl\{L(\psi\!+\delta\psi, \psi^*\! + \delta\psi^*\!,
  \partial_\mu\psi\! + \delta\partial_\mu\psi,
   \partial_\mu\psi^*\! + \delta\partial_\mu\psi^*)
    \nonumber\\
 -L(\psi, \psi^*\!, \partial_\mu\psi, \partial_\mu\psi^* )\bigr\} d(x)
  \hspace{31mm}
\end{eqnarray*}
For simplicity in the above expression, we omitted the
subscripts of $\psi$ labelling different
components, abbreviated $d(x)$ for $d t dx dy dz$ and
$\partial_\mu$ for $\partial/\partial x_\mu $
with $\{ x_\mu\} = \{t, x, y, z\}$.
Because of, explicitly,
$\delta\psi_a=i(\delta\vartheta_a)\psi_a$,
$\delta\psi_b=i(\delta\vartheta_b)\psi_b$,
$\delta\psi_{m (g)}=i(\delta\vartheta_a + \delta\vartheta_b)\psi_{m (g)}$,
and their complex conjugations,
we derive the following result,
\begin{equation}
\delta I = \int\bigl\{
  (\partial_\mu J^\mu_A ) \delta\vartheta_a
  +( \partial_\mu J^\mu_B) \delta\vartheta_b
   \bigr\}d(x),
\end{equation}
where $J^\mu_A = J_a^\mu + J_m^\mu + J_g^\mu$,
$J^\mu_B = J_b^\mu + J_m^\mu + J_g^\mu$
with
\begin{equation}
J^\mu_i =  \psi_i \frac{\delta L}{\delta (\partial_\mu\psi_i)}
  -\psi_i^*\frac{\delta L}{\delta (\partial_\mu\psi_i^*)},
  \quad i=a,\,b,\,m,\,g.
\end{equation}
The system is invariant under the transformation
Eq.~(\ref{eq:trans}) such that $\delta I=0$,
which gives rise to two conserved currents.
\begin{equation}\label{eq:conserved-currents}
\partial_\mu J_A^\mu=0, \quad \partial_\mu J_B^\mu = 0.
\end{equation}

In the present paper, we focus on  uniform system
($\nabla\psi_i \approx 0$) and hence
neglect the kinetic and trapping potential terms.
Then the conservation law Eq.~(\ref{eq:conserved-currents}) reads
$\displaystyle\frac{d}{dt}(|\psi_a|^2+|\psi_m|^2+|\psi_g|^2)=0$,
$\displaystyle\frac{d}{dt}(|\psi_b|^2+|\psi_m|^2+|\psi_g|^2)=0$,
which implies that
\begin{eqnarray}\label{eq:constraints}
|\psi_a|^2+|\psi_m|^2+|\psi_g|^2 &=& n^{}_a,
 \nonumber\\
|\psi_b|^2+|\psi_m|^2+|\psi_g|^2 &=& n^{}_b,
\end{eqnarray}
where the constants $n^{}_a$ and $n^{}_b$ are determined
by $V$, the volume of the system, together with $N_a(0)$ and $N_b(0)$,
the initial numbers  of species $a$ and $b$, \ie,
$n^{}_a=N_a(0)/V$, $n^{}_b=N_b(0)/V$.
Here we assume there are no molecules at the initial time in the system.

To guarantee the compatibility with
the constraints given by Eq.~(\ref{eq:constraints}),
we introduce two Lagrange multipliers
$\mu_a^{}$ and $\mu_b^{}$ into the mean-field Lagrange density
Eq.~(\ref{eq:Lag}),
\begin{equation}\label{eq:LC}
K=L+\mu_a^{}|\psi_a|^2
+\mu_b^{}|\psi_b|^2+(\mu_a^{}+\mu_b^{})(|\psi_m|^2+|\psi_g|^2).
\end{equation}
Here the real parameters $\mu^{}_a$ and $\mu^{}_b$
can be identified as the chemical potentials of the
corresponding components.
Owning to Eq.~(\ref{eq:constraints}), we can conveniently
introduce new notations:
$\phi_i^{}=\psi_i^{}/\sqrt{n_a^{}+n_b^{}}$,
$\lambda_{ij}=\lambda_{ij}'(n_a^{}+n_b^{})/\hbar$,
$\alpha=\alpha'\sqrt{n_a^{}+n_b^{}}/\hbar$,
$A_i=A_i'(n_a^{}+n_b^{})^{2/3}/\hbar$, $\Omega=\Omega'/\hbar$,
$\mathcal{E}=\mathcal{E}'/\hbar$, and $\Delta=\Delta'/\hbar$.
Reexpressing equation~(\ref{eq:LC}) in terms of $\phi$'s and substituting
it into the Euler-Lagrange equation, we obtain a set of equations
\begin{eqnarray} \label{eq:DE}
i\frac{\partial \phi_a^{}}{\partial t}&=&\sum_{i\neq
a}\lambda_{ai} |\phi_i|^2\phi_a^{}+A_a|\phi_a|^{4/3}\phi_a^{}
+\alpha\phi_b^*\phi_m-\mu^{}_a\phi_a^{},
   \nonumber\\
i\frac{\partial \phi_b^{}}{\partial t}&=&\sum_{i\neq
b}\lambda_{bi} |\phi_i|^2\phi_b^{}+A_b|\phi_b|^{4/3}\phi_b^{}
+\alpha\phi_a^*\phi_m-\mu_b^{}\phi_b^{},
   \nonumber\\
i\frac{\partial\phi_m^{}}{\partial
t}&=&\sum_i\lambda_{mi}|\phi_i|^2\phi_m^{}
+\alpha\phi_a^{}\phi_b^{}-\Omega\phi_g^{} +\mathcal{E}\phi_m^{}
  \nonumber\\
&&-i\gamma\phi_m^{}-(\mu_a^{}+\mu_b^{})\phi_m^{},\nonumber\\
i\frac{\partial\phi_g^{}}{\partial
t}&=&\sum_i\lambda_{gi}|\phi_i|^2\phi_g^{} -\Omega\phi_m^{}
+\Delta\phi_g^{}
 \nonumber\\
&&-(\mu_a^{}+\mu_b^{})\phi_g^{},
\end{eqnarray}
where a phenomenological parameter $\gamma$ is introduced  to characterize
the decay of quasibound molecules.
In terms of $\phi_i^{}$'s, the conservation relations
(\ref{eq:constraints})
turn to be
\begin{eqnarray}\label{eq:constraint-redef}
|\phi_a^{}|^2+|\phi_m^{}|^2+|\phi_g^{}|^2=(1+\delta)/2,
 \nonumber\\
|\phi_b^{}|^2+|\phi_m^{}|^2+|\phi_g^{}|^2=(1-\delta)/2,
\end{eqnarray}
with $\delta=(n_a^{}-n_b^{})/(n_a^{}+n_b^{})$ characterizing the
population imbalance between fermionic atoms in different states.

\subsection{The definition of fidelity}

Now we are in the position to introduce a
proper definition of fidelity for our system.
As our system is a four-component system of Bose-Fermi
mixture which is related to a graded unitary group
SU$(2\mid2)$, we need to  define the fidelity
carefully as it must obey several basic properties~\cite{Nie}.
Equation (\ref{eq:constraint-redef}) actually provides us
the normalization condition
\[
|\phi_a|^2 + |\phi_b|^2 + 2|\phi_m|^2  + 2|\phi_g|^2 = 1,
\]
which can be expressed as the following form
\begin{equation}\label{eq:normalization}
\bra{\phi}F^*(\phi)F(\phi)\ket{\phi}=1,
\end{equation}
where $\bra{\phi}$ denotes $(\phi^*_a, \phi^*_b, \phi^*_m,
\phi^*_g)$. One might think of a naive expression for the
$F$-matrix, $F=\mathrm{diag}(1, 1, \sqrt{2}, \sqrt{2})$. However,
because the relation of Eq.~(\ref{eq:normalization}) should be
invariant under the transformation given in Eq.~(\ref{eq:trans}),
just like that the conventional inner product in quantum mechanics is
invariant under the U(1) transformation, the simplest correct
expression of the $F$-matrix ought to be
\begin{eqnarray}\label{eq:F}
F(\phi)=\left(
\begin{array}{cccc}
  \displaystyle\frac{\phi_b}{|\phi_b|} & 0 & 0 & 0 \\
  0 & \displaystyle\frac{\phi_a}{|\phi_a|} & 0 & 0 \\
  0 & 0      &           \sqrt{2} & 0 \\
  0 & 0      &  0   &       \sqrt{2} \\
\end{array}%
\right).
\end{eqnarray}
As a result, a nature definition of fidelity of a state labelled by
$\phi$ with that labelled by $\phi^\prime$ is given by
\begin{equation}\label{eq:fidelity}
f(\phi, \phi^\prime)=\mid\bra{\phi}F^*(\phi)F(\phi^\prime)\ket{\phi^\prime}\mid,
\end{equation}
where the $F$-matrix was given in Eq.~(\ref{eq:F}).
Clearly, such a definition fulfills
$f(\phi, U(\vartheta_a,\vartheta^{}_b )\phi^\prime)
=f(U(\vartheta_a,\vartheta^{}_b )\phi, \phi^\prime)=f(\phi, \phi^\prime)$,
which means the phase transformation given in Eq.~(\ref{eq:trans})
does not vary the magnitude of fidelity; and
the fidelity of a state with itself is always the unit
$f(\phi, \phi)=1$
which is just the normalization condition .

\section{Coherent population trapping states}\label{sec:cpt-state}

Now we consider the stationary states where
we neglect the decay of quasibound molecules (\ie, $\gamma=0$). We
know that the existence of stationary solutions of Eq.(
\ref{eq:DE}) requires the system satisfies the adiabatic
approximation condition. Once the adiabatic approximation is
valid, \ie, $\partial{\phi_i^{}}/\partial t\approx 0$, the
time-evolution equations (\ref{eq:DE}) become algebraic ones for
$\phi$'s, namely,
\begin{eqnarray} \label{eq:station}
\mu^{}_a\phi_a^{}&=&\sum_{i\neq a}\lambda_{ai}
|\phi_i|^2\phi_a^{}+A_a|\phi_a|^{4/3}\phi_a^{}
+\alpha\phi_b^*\phi_m,
   \nonumber\\
\mu_b^{}\phi_b^{}&=&\sum_{i\neq b}\lambda_{bi}
|\phi_i|^2\phi_b^{}+A_b|\phi_b|^{4/3}\phi_b^{}
+\alpha\phi_a^*\phi_m,
   \nonumber\\
(\mu_a^{}+\mu_b^{})\phi_m^{}&=&\sum_i\lambda_{mi}|\phi_i|^2\phi_m^{}
+\alpha\phi_a^{}\phi_b^{}-\Omega\phi_g^{}
+\mathcal{E}\phi_m^{},
    \nonumber\\
(\mu_a^{}+\mu_b^{})\phi_g^{}&=&\sum_i\lambda_{gi}|\phi_i|^2\phi_g^{}
-\Omega\phi_m^{}
+\Delta\phi_g^{}.
\end{eqnarray}
Although it is difficult to find the exact solutions of the above
equations, one can easily obtain a set of steady state solutions for
Eq.~(\ref{eq:station}) by taking  $\phi_m^{}=0$.
Such a state is called coherent population trapping (CPT) state which
yields,
\begin{eqnarray} \label{eq:cpt}
|\phi_a^0|^2&=&\frac{(\delta-\tilde{\Omega}^2)
+\sqrt{(\delta-\tilde{\Omega}^2)^2+2\tilde{\Omega}^2(1+\delta)}}{2},
\nonumber\\
|\phi_b^0|^2&=&|\phi_a^0|^2-\delta,\quad\quad\quad
|\phi_g^0|^2=\frac{1+\delta}{2}-|\phi_a^0|^2,\nonumber\\
\mu_a^{}&=&\lambda_{ab}|\phi_b^0|^2+\lambda_{ag}|\phi_g^0|^2
+A_a|\phi_a^0|^{4/3},\nonumber\\
\mu_b^{}&=&\lambda_{ab}|\phi_a^0|^2+\lambda_{bg}|\phi_g^0|^2
+A_b|\phi_b^0|^{4/3},
\end{eqnarray}
where $\tilde{\Omega}=\Omega/\alpha$. The resonance condition
corresponding to this solution is
\begin{widetext}
\begin{equation}\label{eq:RC}
\Delta=(\lambda_{ab}-\lambda_{ag})|\phi_a^0|^2+
(\lambda_{ab}-\lambda_{bg})|\phi_b^0|^2+(\lambda_{ag}+\lambda_{bg}
-\lambda_{gg})|\phi_g^0|^2+A_a|\phi_a^0|^{4/3}+A_b|\phi_b^0|^{4/3}.
\end{equation}
\end{widetext}
We consider a laser pulse with $\tilde{\Omega}\rightarrow\infty$
for $t\rightarrow 0$ and $\tilde{\Omega}\rightarrow 0$ for $t\rightarrow\infty$.
For such a laser field, at the initial time ($t=0$),
we have $|\phi_a^0|^2=(1+\delta)/2$,
$|\phi_b^0|^2=1-\delta/2$ and $|\phi_g^0|^2=0$,
which implies there are no molecules in the system at the initial time.
At the final time ($t\rightarrow\infty$),
$|\phi_a^0|^2\rightarrow\delta$,
$|\phi_b|^2\rightarrow 0$ and
$|\phi_g^0|^2\rightarrow(1-\delta)/2$ for $\delta>0$; whereas
$|\phi_a^0|^2\rightarrow 0$, $|\phi_b|^2\rightarrow -\delta$ and
$|\phi_g^0|^2\rightarrow(1+\delta)/2$ for $\delta<0$.
With the help of initial values of $|\phi_i^0|^2$
and their asymptotic values at final time,
it is easy to find that those fermionic atoms,
in the presence of their counterparts,
can be converted into molecules if the CPT states
can be followed adiabatically.
The residual atoms can not be converted into molecules due to the lack of
counterpart atoms.
After the numerical calculation in next section,
we will go back to study whether the CPT state can be followed adiabatically
with the help of the useful concept, adiabatic fidelity.

\section{On conversion efficiencies}\label{sec:efficiency}

We know that there are two sorts of fermionic atoms $^6$Li
and $^{40}$K in the group of alkali-metal atoms.
They have been converted  into molecules in experiments successfully
through the bare Feshbach resonance~\cite{Regal,Stre, JC,Jin}.
In those experiments, 60\% to 80\%  of $^{40}$K atoms and no more than
85\% of $^6$Li atoms can be converted into molecules.
For $^6$Li atoms, the atom-to-molecule conversion efficiency via
bare Feshbach resonance is lower than that via STIRAP technique aided by Feshbach
resonance, which is in contrast to the case for $^{40}$K atoms.
Due to the difference in atomic properties,
the atom-to-molecule conversion efficiency differs for different atoms
even if the same technique is applied.

Now we evaluate the atom-to-molecule conversion efficiency,
respectively, for $^6$Li and $^{40}$K atoms with concrete magnetic
and laser fields.
For these two sorts of atoms, we adopt the same time-dependent Rabi
frequency,
\begin{equation}
\Omega(t)=\Omega_{max}\Bigl[1-\tanh\Bigl(\frac{t-t_0}{\tau}\Bigr)\Bigr],
\end{equation}
where the parameters $\Omega_{\mathrm{max}}$, $t_0$ and $\tau$
are determined by the applied laser field
that couples the two molecular states.
In the numerical calculation, the detuning
strength $\Delta$ is given by Eq.~(\ref{eq:RC}).
We assume there are no molecules in the system at the initial time
\ie, $\phi_{m,g}=0$ at $t=0$.
\begin{figure}[h]
\vspace{-7mm}
\includegraphics[width=75mm]{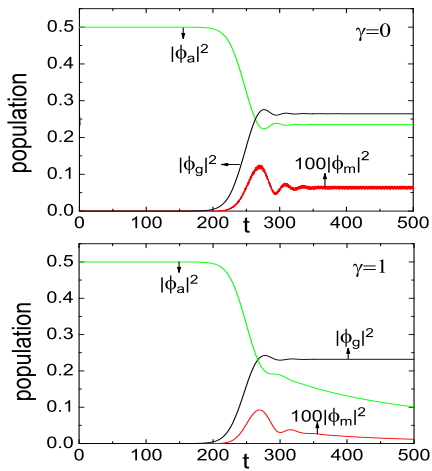}
\vspace{-10mm}
 \caption{\label{fig:K}  (color online)
 The time dependence of the population of particles for  $^{40}$K system
 for different $\gamma$. The parameters are $\delta=0$, $\lambda^{}_{ab}=0.24$,
$A_a^{}=A_b^{}=0.16$, $\mathcal{E}=-4.4$,
$\Omega_{\mathrm{max}}=200$, $t_0=120$, $\tau=40$, and the other
parameters are zero. Time is in unit of $1/\alpha$ and all other
coefficients  are in units of $\alpha$, where
$\alpha=16.6\times10^{-29}$J.}
\end{figure}

For $^{40}$K atoms, we know that the Feshbach resonance occurs at
a magnetic field strength of 202.1G and the resonance width is
about 7.8G~\cite{Re}.
Then we can get the atom-to-molecule coupling
strength $\alpha'=16.6 \times 10^{-39}\mathrm{J}$
according to Ref.~\cite{Lu}.
Here we choose that the magnetic field is $201.7$G and
particle density $n_a+n_b$ is about $10^{20}$m$^{-3}$.
It is easy
to obtain $A_a=0.16\alpha$, $\lambda_{ab}=0.24\alpha$ and
$\mathcal{E}=-4.4\alpha$.
The time evolution of the corresponding population
can be obtained by solving Eq.~(\ref{eq:DE}).
We plot the numerical results in Fig.~\ref{fig:K}.
From this figure, we can see that the conversion efficiency
$2|\phi_g(t=\infty)|^2$ for $^{40}$K atoms is less than
60\%, lower than that via the technique of bare Feshbach
resonance~\cite{Jin}.
The low conversion efficiency implies that the
CPT state can not be followed adiabatically, which will be  confirmed
confidently by evaluating the adiabatic fidelity in next section.
Comparing the two panels in Fig.\ref{fig:K},
we can know that the influence of $\gamma$ on the atom-to-molecule
conversion efficiency is very small. This is due to that $|\phi_m|^2$ is
close to zero at any time in contrast to the case for $^6$Li atoms.
\begin{figure}[h]
\vspace{-7mm}
\includegraphics[width=75mm]{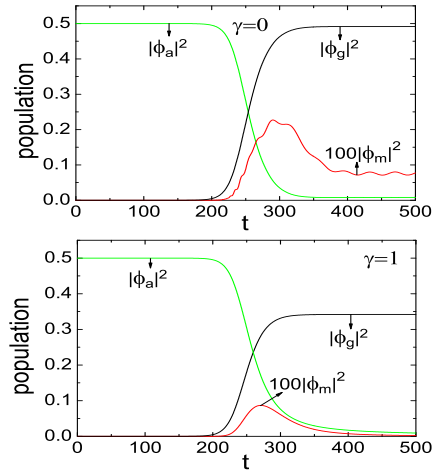}
\vspace{-10mm}
 \caption{\label{fig:Li}(color online) The time dependence of the population
 of particles
 for  $^6$Li system for different $\gamma$.
 The parameters are $\delta=0$, $\lambda^{}_{ab}=0.0027$,
$A_a^{}=A_b^{}=0.055$, $\mathcal{E}=-0.1125$,
$\Omega_{\mathrm{max}}=200$, $t_0=120$, $\tau=40$, and the other
parameters are zero. Time is in unit of $1/\alpha$ and all other
coefficients  are in units of $\alpha$, where
$\alpha=3.29\times10^{-27}$J.}
\end{figure}

For $^6$Li atoms, the Feshbach resonance occurs
at two distinct strengths of magnetic field
(the called narrow and broaden Feshbach resonance respectively).
In our calculation, we focus on the narrow Feshbach resonance
for which the atom-to-molecule coupling strength
$\alpha'$ is about 3.29$\times10^{-37}$J.
If the particle density $n_a+n_b$ is about
$10^{20}$m$^{-3}$ and the magnetic field is about 543.6G, one can
get $A_a=0.055\alpha$, $\lambda_{ab}'=0.0027\alpha$ and
$\mathcal{E}=-0.1125\alpha$.
Figure \ref{fig:Li} shows the time evolution of particle
populations for the conversion of $^6$Li atoms into molecules.
From the top panel, we find that almost
all of $^6$Li atoms can be converted into molecules.
For $^6$Li atoms, the atom-to-molecule conversion efficiency
via the STIRAP technique aided by Feshbach resonance is higher
than that via bare Feshbach resonance~\cite{Stre,JC},
which is in contrast to the case for $^{40}$K atoms.
From the bottom panel, we can see that the atom-to-molecule
conversion efficiency decreases distinctly
due to the  existence of the quasibound molecular decay.

\begin{figure}[h]
\includegraphics[width=75mm]{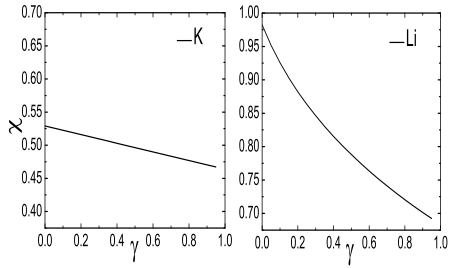}
\vspace{-7mm}
 \caption{\label{fig:gama}
 The atom-to-molecule conversion efficiency versus
 the decay rate of the quansibound  molecules for
 $^{40}$K (left panel) and for $^6$Li (right panel) system.
 The parameter  choice for left (right) panel is the same as Fig.~\ref{fig:K}
 (Fig.~\ref{fig:Li}). }
\end{figure}

The dependence of the conversion efficiency on the decay rate $\gamma$
of quasibound molecules
for both $^{40}$K and $^6$Li is plotted in Fig.~\ref{fig:gama}.
Clearly,
the influence of the decay on the conversion efficiency
for $^6$Li atoms is more distinct than that for $^{40}$K atoms,
which confirms the previous interpretation.

\section{Adiabatic fidelity for CPT states}\label{sec:fidelity}

From the above numerical results, we can find that the
atom-to-molecule conversion efficiencies of
STIRAP technique aided by Feshbach resonance for $^{40}$K and $^6$Li are
distinctly different although the CPT states are assumed to
exist for the two systems.
As we know that most of the atoms can be converted into molecules
only when the CPT state is followed adiabatically, and
the existence of the CPT state does not
guarantee the state can be followed adiabatically~\cite{Ling}.
The adiabatic properties for the atom-to-molecule conversion
have been studied in Ref.~\cite{Matt}.
Whereas, we can not apply this method to our system
since all the possible solutions of the stationary equation (\ref{eq:DE})
must be known in the approach~\cite{Matt}
and it is difficult to obtain the other solutions
beyond the CPT-state solution in our system.
\begin{figure}[h]
\includegraphics[width=85mm]{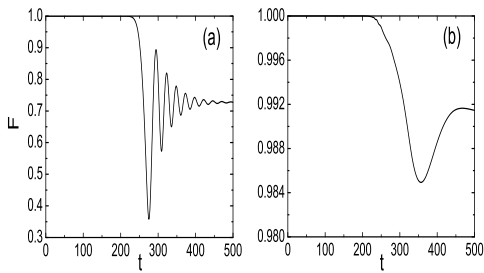}
\vspace{-7mm}
 \caption{\label{fig:fidelity}
 The time dependence of adiabatic fidelity for CPT state:
 ({\bf a}) for $^{40}$K, the same parameter choice as in Fig.\ref{fig:K};
 ({\bf b}) for $^{6}$Li, the same parameter choice as in Fig.\ref{fig:Li}.
 }
\end{figure}

Recently, the fidelity was employed~\cite{Ying} to characterize the adiabatic
condition for systems with atom-to-molecule conversion.
The key point is that the value of the fidelity should close to unity
if the system can adiabatically evolve in the CPT state.
Now we evaluate the adiabatic fidelity $f(\phi^0(t),\phi(t))$ for
our model, in which $\phi(t)$ is the wave function obtained by
solving the dynamical Eq.~(\ref{eq:DE}) with $\gamma=0$, and
$\phi^0(t)$ is the wave function corresponding to the CPT state
which is given in Eq.~(\ref{eq:cpt}).
The time evolution of the
adiabatic fidelity for CPT state is plotted in
Fig.~\ref{fig:fidelity}(a) for $^{40}$K system.
One can see that the magnitude of fidelity is about $1$
at the initial time, which implies the system adiabatically
evolves along CPT state. At the time $275/\alpha$, the fidelity
diminishes to the minimal value $0.35$, which implies the system
deviates away from the CPT state distinctly at that time. Although the
fidelity begins to fluctuate later on, its final value is still no
more than $0.75$.
The above analysis implies that the CPT state can
not be followed adiabatically, so the atom-to-molecule conversion
efficiency is not high for $^{40}$K atoms.
Fig.~\ref{fig:fidelity}(b) shows the time evolution of the
adiabatic fidelity for CPT state corresponding to the conversion
of $^6$Li atoms into molecules. Obviously, the fidelity for this
system is very close to $1$ at all the time, which implies that the system
adiabatically evolves along the CPT state. This result is
consistent with the fact that almost all of the $^6$Li atoms can
be converted into molecules.
Since the fidelity for CPT state can well reflect
the atom-to-molecule conversion efficiency, one can improve the
conversion efficiency by optimizing the parameters of the
system to achieve a higher adiabatic fidelity.

\section{Summary}\label{sec:summary}

With the help of mean-field Lagrangian density,
we studied the conversion of two-species fermionic atoms into bosonic
molecules via STIRAP technique aided by Fechbach resonance.
We calculated conversion efficiencies for $^{40}$K and $^6$Li systems
respectively, and found that almost all of the $^6$Li atoms
can be converted into molecules,
which implies that the STIRAP technique aided by Feshbach resonance
is more effective than the bare Fechbach resonance for $^6$Li atoms
rather than $^{40}$K atoms.
We also compared the influence of the decay rate of quansibound molecules on
the conversion efficiency for $^{40}$K and $^6$Li systems respectively,
and found that there is a big difference between them.
The success of STIRAP technique not only requires the
existence of the CPT state but also requires that the system can
adiabatically evolve within such a state.

The adiabatic fidelity was recently recognized to be a useful measurement
for characterizing the adiabatic properties.
Our analysis of the symmetry and the corresponding conservation law
for the systems under consideration helped us to introduce
an appropriate definition of adiabatic fidelity for CPT state.
For $^{40}$K system, the CPT state can not be adiabatically followed
since the fidelity was found to be less than $0.75$ at final time,
which is well consistent with the corresponding low conversion efficiency.
In order to improve the conversion efficiency in $^{40}$K system,
one should achieve a higher fidelity through optimizing
the parameters of the system.
Whereas, for $^6$Li system, the fidelity is very close to $1$,
hence the CPT state can be followed adiabatically,
which is the reason for a high conversion efficiency.
Our evaluation of the adiabatic fidelity
enable us to understand why the conversion efficiencies for $^{40}$K and $^6$Li
are distinctly different.

The work was supported by Program for Changjiang Scholars and Innovative
Research Team in University, and NSFC Grant No. 10674117.


\begin{thebibliography}{99}

\bibitem{Joch}
S. Jochim, M. Bartenstein, A. Altmeyer, G. Hendl, S. Riedl, C.
Chin, J. H. Denschlag, and  R. Grimm, Scince {\bf302}, 2101
(2003).

\bibitem{zwi}
M. W. Zwierlein, C. A. Stan, C. H. Schunck, S.M. F. Raupach, S.
Gupta, Z. Hadzibabic, and W. Ketterle, Phys. Rev. Lett. {\bf91},
250401 (2003).

\bibitem{Greiner}
M. Greiner, C. A. Regal and  D. S. Jin, Nature {\bf426}, 537
(2003).

\bibitem{Matt}
M. Mackie, R. Kowalski and J. Javanainen, Phys. Rev. Lett.
{\bf84}, 3803 (2000).

\bibitem{Alze}
G. Alzetta et al., Nuovo Cimento Soc. Ital. Fis., B 36, 5 (1976);
G. Alzetta, L. Moi, and G. Orriols, ibid. 52, 209 (1979).

\bibitem{Ying}
S. Y. Meng, L. B. Fu and J. Liu, e-print arXiv:0709.0359.

\bibitem{Regal}
C. A. Regal, C. Ticknor, J. L. Bohn, D. S. Jin, Nature 424, 47
(2003).

\bibitem{Stre}
K. E. Strecker, G. B. Partridge, R. G.
Hulet, Phys. Rev. Lett. 91, 080406 (2003).

\bibitem{JC}
J. Cubizolles, T. Bourdel, S. J. J. M. F. Kokkelmans, G. V.
Shlyapnikov, and C. Salomon, Phys. Rev. Lett. {\bf 91}, 240401
(2003).
\bibitem{Jin}
C. A. Regal, M. Greiner, and D. S. Jin, Phys. Rev. Lett. {\bf 92},
040403 (2004).
\bibitem{Stan}
C. A. Stan, M.W. Zwierlein, C. H. Schunck, S.M. F. Raupach, and W.
Ketterle, Phys. Rev. Lett. {\bf 93}, 143001 (2004).

\bibitem{Ino}
S. Inouye, J. Goldwin, M. L. Olsen, C. Ticknor, J. L. Bohn, and D.
S. Jin, Phys. Rev. Lett. {\bf93}, 183201 (2004).

\bibitem{Lu}
L. H. Lu and Y. Q. Li, Phys. Rev. A {\bf 76}, 053608 (2007).

\bibitem{Re}
C. A. Regal, C. Ticknor, J. L. Bohn, and D. S. Jin, Phys. Rev.
Lett. {\bf90}, 053201 (2003).

\bibitem{Ling}
H. Y. Ling, H. Pu and B. Seaman, Phys. Rev. Lett. {\bf93}, 250403
(2004).

\bibitem{Nie}
M. A. Nielsen and I. L. Chuang, Quantum Computation and Quantum
Information (Cambridge University Press, Cambridge, England,
2000), pp. 399-424.

\end{thebibliography}
\end{document}